\documentclass[aps,preprint,amsmath,amssymb]{revtex4-1}
\usepackage{graphicx}
\usepackage{color}

\def\g5{\gamma_{5}}
\def\ga{\gamma}

\def\be{\begin{eqnarray}}
\def\ed{\end{eqnarray}}

\def\non{\nonumber}

\def\la{\langle}
\def\ra{\rangle}

\begin{document}
\title{ \Large \bf
Z-mediated charge and CP asymmetries\\ and FCNCs in $B_{d,s}$ processes
}
\date{\today}

\author{ \bf  Chuan-Hung Chen$^{1,2}$\footnote{Email:
physchen@mail.ncku.edu.tw}, Chao-Qiang Geng$^{3}$\footnote{Email:
geng@phys.nthu.edu.tw} and Wei Wang$^{4}$\footnote{Email:
wei.wang@ba.infn.it}
 }

\affiliation{ $^{1}$Department of Physics, National Cheng-Kung
University, Tainan 701, Taiwan \\
$^{2}$National Center for Theoretical Sciences, Hsinchu 300, Taiwan\\
$^{3}$Department of Physics, National Tsing-Hua University, Hsinchu
300, Taiwan \\
$^{4}$Istituto Nazionale di Fisica Nucleare, Sezione di Bari, Bari
70126, Italy
 }

 \begin{abstract}
We  show model-independently that the negative like-sign charge asymmetry  $(-A^b_{s\ell})$  is less than $ 3.16\times 10^{-3}$ when the constraints from the $B_q-\bar B_q$ mixings and the time-dependent CP asymmetries (CPAs) for $B_q\to J/\Psi M_q$ with $M_q=K,\phi$ and $q=d,s$ are taken into account. Although the result is smaller than the measured value by the  D{\O} Collaboration at Fermilab, there is still plenty of room to have new physics, which is sensitive to new CP violating effects, as the standard model (SM) prediction is $(2.3_{-0.5}^{+0.6})\times 10^{-4}$. To illustrate the potential large $|A^{b}_{s\ell}|$, we show the influence of new $SU(2)_L$ singlet exotic quarks in the vector-like quark model, where  the $Z$-mediated flavor changing neutral currents (FCNCs) are generated at tree level. In particular, we demonstrate that (a) the like-sign charge asymmetry could be enhanced by a factor of two in magnitude; (b) the CPA of $\sin2\beta^{J/\Psi \phi}_s$ could reach to $-15\%$; (c) the CPA of $\sin2\beta_{\phi K_S}$ could be higher than $\sin2\beta_{J/\Psi K_S}$ when $|A^b_{s\ell}|$ is larger than the SM prediction; and (d) the branching ratio for $B_s\to \mu^+ \mu^-$ could be as large as $0.6\times 10^{-8}$.

 \end{abstract}
 \maketitle

\section{Introduction}

It is clear that some new CP violation mechanism beyond the Kobayashi-Maskawa (KM) phase in the standard model (SM) is needed in order to explain the matter-antimatter asymmetry of the Universe. Moreover, several hints for the existence of some new CP violating phases are revealed in the low energy processes,  such as the $\pi K$ puzzle in $B\to \pi K$ decays, the large CP asymmetry (CPA) of $\sin2\beta^{J/\Psi \phi}_s$ in the $B_s\to J/\Psi \phi$ decay, inconsistent time-dependent CPAs between $B_d\to (\eta, \phi) K_S$ and $B_d\to J/\Psi K_S$ decays, etc \cite{Barberio:2008fa}.

Recently, the D{\O} Collaboration at Fermilab has observed the like-sign charge asymmetry, defined as \cite{Abazov:2010hv}
 \be
A^{b}_{s\ell} &=& \frac{N^{++}_b - N^{--}_{b}}{N^{++}_{b}+N^{--}_{b}}\,, \label{eq:Absl_exp}
 \ed
where $N^{++(--)}_b$ denotes the number of events that $b$ and $\bar b$-hadron semileptonically decay into two positive (negative) muons. The measured value in the dimuon events is given by
 \cite{Abazov:2010hv}
 \be
 A^{b}_{s\ell} = \left(-9.57 \pm 2.51({\rm stat}) \pm 1.46 ({\rm syst}) \right)\times 10^{-3}\,, \label{eq:d0}
 \ed
which is about 3.2 standard deviations from the SM prediction of
$(-2.3^{+0.5}_{-0.6})\times 10^{-4}$ \cite{Abazov:2010hv,Lenz:2006hd}. If the semileptonic b-hadron decays do not involve a CP phase, the charge asymmetry is directly related to the mixing-induced CPAs in $B_{d,s}$-meson oscillations (see the detailed analysis later). 
Although the errors of the data are still large, the 
deviations from the SM could be attributed to the new CP violating phases in $b\to d$ and $b\to s$ transitions \cite{Randall:1998te,Dighe:2010nj,Dobrescu:2010rh,Chen:2010wv,Buras:2010mh,Ligeti:2010ia,Bauer:2010dg,Deshpande:2010hy,Choudhury:2010ya,Lenz:2007nj,Bobrowski:2009ng,Buchalla:2000sk,Altmannshofer:2009ne}.

Inspired by the new D{\O} measurement and other CPAs measured earlier, we illustrate that the anomalies can be induced by the  new exotic vector-like quarks in the so-called vector-like-quark model (VQM). Unlike the conventional four-generation model with the fourth left-handed quarks being an $SU(2)_L$ doublet, the vector-like quarks (VQs) are all $SU(2)_L$ singlets, as the ones naturally realized in $E_6$ models \cite{E_6}. Since the left-handed VQs carry the same hypercharge as the right-handed quarks in the SM, interestingly  the model leads to Z-mediated flavor changing neutral currents (FCNCs) at tree level \cite{ZFCNC,Chen:2008ug,Higuchi:2009dp,Barenboim:1997qx}.
Moreover, the VQM involves less free parameters and is more predictable since the couplings of Z-boson to fermions and $m_Z$ are known. In addition to the mixing-induced CPAs, the VQM  has  significant impacts on the rare $B_q$ decays such as  $b\to s \ell^{+} \ell^{-}$ and $B_s\to \mu^{+} \mu^{-}$ as well as  other $B_q$ processes.

The paper is organized as follows. In Sec.~\ref{sec:wrong-like}, we   analyze model-independently
the wrong and like-sign charge asymmetries in detail. 
In Sec.~\ref{sec:model}, we derive Feynman rules for the Z-mediated FCNCs in the VQM and formulate  CPAs and rare $B_q$ decays. The numerical analysis 
is presented in Sec.~\ref{sec:num}. The conclusion is given in Sec.~\ref{sec:conclusion}

\section{Model-independent results on charge asymmetries }\label{sec:wrong-like}

In order to comprehend the implication of the like-sign charge asymmetry $A^{b}_{s\ell}$, we use 
both  experimental   and phenomenological approaches. We first discuss the issue from the viewpoint of 
the current data. To evaluate $A^{b}_{s\ell}$,
we start with the wrong-sign charge asymmetry in semileptonic $B_q$ decays,
 defined by \cite{PDG08}
 \be
a^q_{s\ell}&=& \frac{\Gamma(\bar B_q(t) \to \ell^+ X)- \Gamma( B_q(t) \to \ell^- X)}{\Gamma(\bar B_q(t) \to \ell^+ X)
+\Gamma( B_q(t) \to \ell^- X)}\,,\non \\
&\approx& Im\left( \frac{\Gamma^{q}_{12}}{M^{q}_{12}}\right) \label{eq:aqsl}
 \ed
where $\Gamma^{q}_{12}(M^q_{12})$ denotes the absorptive (dissipative) part of the $B_q
\leftrightarrow \bar B_q$ transition with $\Gamma^q_{12}\ll M^q_{12}$.
As a consequence, a non-zero $a^q_{s\ell}$ indicates  CP violation.  
As   $\Gamma^{q}_{12}$ is dominated by the SM contributions,  we adopt $\Gamma^{q}_{12}=\Gamma^{q}_{12}({\rm SM})$ in the following analysis. The SM predictions are $a^d_{s\ell}(\rm SM)=(-4.8^{+1.0}_{-1.2})\times 10^{-4}$ and $a^{s}_{s\ell}(\rm SM)=(2.06\pm 0.57)\times 10^{-5}$ \cite{Lenz:2006hd}, 
while the current data are  $a^d_{s\ell}(\rm Exp)=(-4.7\pm 4.6)\times 10^{-3}$ \cite{Barberio:2008fa} and $a^{s}_{s\ell}(\rm Exp)=(-1.7\pm 9.1)\times 10^{-3}$ \cite{Abazov:2009wg}. The relation between the wrong and like-sign charge asymmetries  indeed can  be expressed by~\cite{Abazov:2010hv,
Grossman:2006ce}
 \be
 A^b_{s\ell} &=& \frac{\Gamma(b\bar b\to \ell^+ \ell^+ X) - \Gamma(b\bar b\to
 \ell^- \ell^- X)}{\Gamma(b\bar b\to \ell^+ \ell^+ X) + \Gamma(b\bar b\to \ell^- \ell^- X)}\,,\non\\
 &=& 0.506(43) a^d_{s\ell} + 0.494(43) a^s_{s\ell}\,. \label{eq:Absl}
 \ed
  From Eq.~(\ref{eq:Absl}), it is easy to see that the like-sign charge asymmetry  depends on the CP phases in $B_d$ and $B_s$  oscillations. If we take $a^d_{s\ell}(\rm Exp)$ and the D{\O} observed value of $A^{b}_{s\ell}$ as inputs, we immediately get
 \be
A^{s}_{s\ell} = 0.494(43) a^s_{s\ell} = (-7.2\pm 3.7)\times 10^{-3}\,.
 \ed
In other words, the wrong-sign charge asymmetry $a^s_{s\ell}$ can be extracted as
 \be
a^{s}_{s\ell}(\rm Extr) = -0.01456\pm 0.00764\,,
 \ed
where 
 the errors have been regarded as uncorrelated and combined in quadrature. Similarly, if $a^{d}_{s\ell}$ is negligible, $a^{s}_{s\ell}(\rm Extr)=-0.01937 \pm 0.0061$. Clearly, by the
 current experimental values,  $|a^s_{s\ell}(\rm Extr)|$ is three orders of magnitude larger than the SM prediction.

\begin{figure}[hpbt]
\includegraphics*[width=5.5 in]{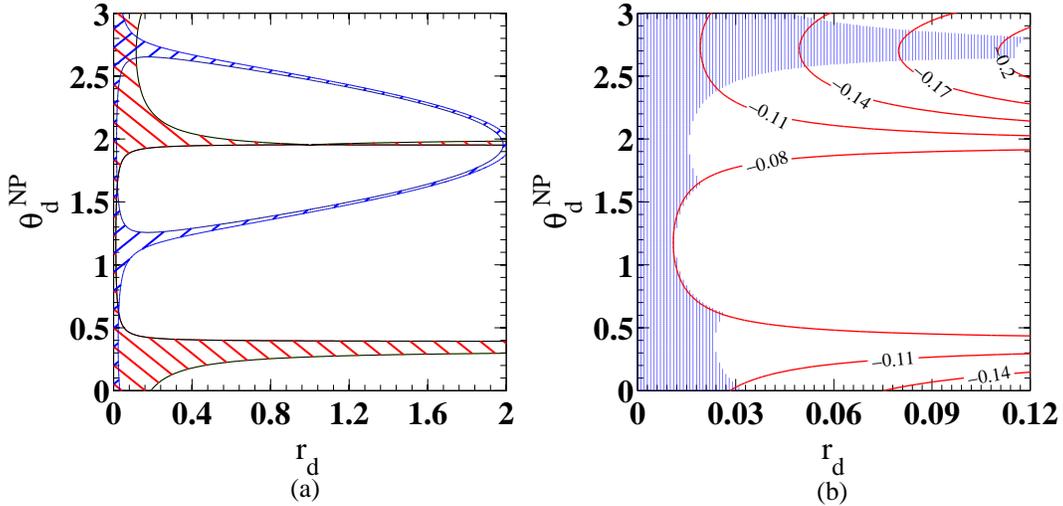}
\caption{ (a) Constraints from 2$\sigma$ errors of $(\Delta m_{B_d})^{\rm Exp}$ (down-left hatched) and $S^{\rm Exp}_{J/\Psi K_S}$ (down-right hatched) and (b) Contours for $\sin(\phi^{\rm NP}_{d} + \phi^{\rm SM}_{d})$ as a function of $r_d$ and $\theta^{\rm NP}_{d}$. }
 \label{fig:limit_d}
\end{figure}

\begin{figure}[hpbt]
\includegraphics*[width=5.5 in]{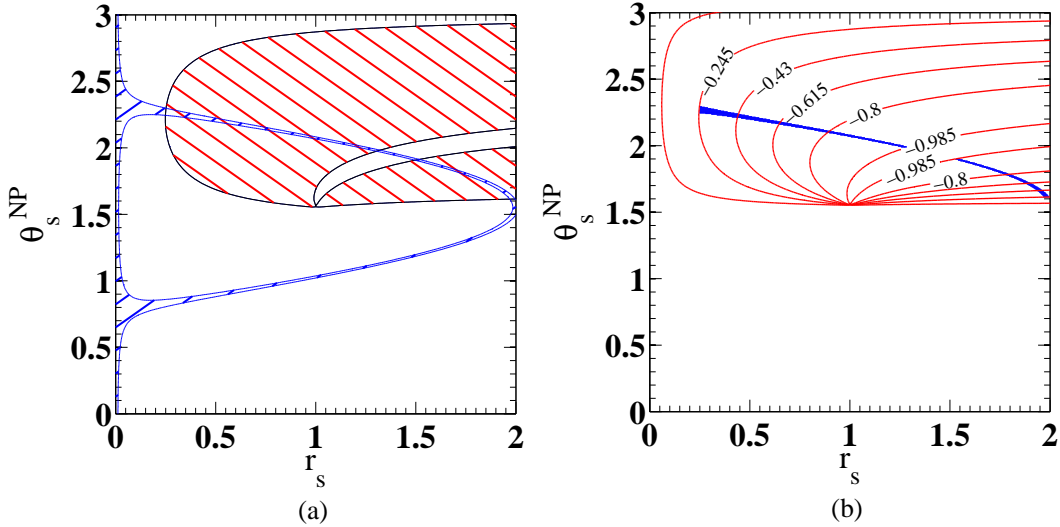}
\caption{ Legend is the same as Fig.~\ref{fig:limit_d} but for q=s. }
 \label{fig:limit_s}
\end{figure}

After discussing the allowed value of $a^{q}_{s\ell}$ from the viewpoint of the current experimental data, it is interesting to analyze the same wrong-sign charge asymmetry from Eq.~(\ref{eq:aqsl}) directly. We set $\Gamma^q_{12}=\Gamma^{q}_{12}(\rm SM)=-|\Gamma^q_{12}(\rm SM)| e^{i\phi^{\Gamma}_{q}}$ and write the $B_q-\bar B_q$ transition matrix element as
 \be
 M^q_{12}&=& M^{q}_{12}( SM) + M^q_{12}( NP)\,, \non \\
 &=& |M^{q}_{12}( SM)| R_q \exp(2i\beta_q + i \phi^{ NP}_{q}) \label{eq:m12_SM_NP}
 \ed
where
 \be
R_q &=& \left(1+r^2_q +2r_q \cos2(\theta^{ NP}_{q}-\beta_q)\right)^{1/2}\,, \non \\
r_{q} &=& \frac{|M^q_{12}( NP)|}{|M^q_{12}( SM)|}\,, \non\\
2\beta_q & =& {\rm arg}(M^q_{12}({ SM}))\,,  \ 2\theta^{ NP}_{q} = {\rm arg}(M^q_{12}({ NP}))\,, \non\\
 \tan\phi^{ NP}_{q} &=& \frac{r_q \sin2(\theta^{ NP}_q -\beta_q)}{1+r_q \cos2(\theta^{ NP}_{q} -\beta_q)} \,. \label{eq:para}
  \ed
With $\Delta\Gamma^q=2|\Gamma^q_{12}|\cos\phi_q$ and $\phi_q= arg\left(-M^q_{12}/\Gamma^{q}_{12} \right)$,  Eq.~(\ref{eq:aqsl}) could be expressed as
 \be
a^q_{s\ell} &\approx& Im\left(\frac{\Gamma^q_{12}}{M^q_{12}} \right) \approx  \frac{\Delta \Gamma^q (\rm SM)}{\Delta m_{B_q}}\frac{\sin\left(\phi^{\rm NP}_q +\phi^{\rm SM}_{q} \right)}{\cos\phi^{\rm SM}_{q}}\,.
\label{eq:aqsl_low}
 \ed
By using the SM results \cite{Lenz:2006hd}:
 \be
\Delta \Gamma_d(\rm SM) &=&(2.67^{+0.58}_{-0.65})\times 10^{-3}\  {\rm ps^{-1} }\,, \non\\
\Delta\Gamma_s (\rm SM) &=&0.096 \pm 0.039\  {\rm ps^{-1} }\,, \non \\
\phi^{\rm SM}_d &=& -0.091^{+0.026}_{-0.038}\,, \non\\
\phi^{\rm SM}_s &=& (4.3 \pm 1.4) \times 10^{-3}\,,
 \ed
and the data: $\Delta m_{B_d}=0.508 \pm 0.005$ ps$^{-1}$ and $\Delta m_{B_s}=17.77 \pm 0.12$ ps$^{-1}$ \cite{PDG08}, 
 we obtain
 \be
 a^d_{s\ell} &=&  (5.26^{+1.14}_{-1.28})\times 10^{-3}\sin(\phi^{\rm NP}_{d} + \phi^{\rm SM}_d)\,, \non \\
 a^s_{s\ell} &=&  (5.40 \pm 2.20)\times 10^{-3}\sin(\phi^{\rm NP}_{s} + \phi^{\rm SM}_s)\,. \label{eq:aqsl_value}
 \ed
Obviously, the sign and magnitude of $a^q_{s\ell}$ are dictated by the factor 
of $\sin\left(\phi^{\rm NP}_q +\phi^{\rm SM}_{q} \right)$. We find that the most strict model-independent constraints on  $\sin(\phi^{\rm NP}_{q} + \phi^{\rm SM}_q)$ are from $\Delta m_{B_q}$ and the time-dependent CPA of
$S_{J/\Psi M_q}=\sin(2\beta_q + \phi^{\rm NP}_{q})$ \cite{PDG08,Chen:2010wv} for $B_q\to J/\Psi M_q$ with $M_q=K(\phi)$ and $q=d(s)$. We note that since $B_q\to J/\Psi M_q$ is dominated by tree diagrams in the SM, 
we have assumed that the contribution to the decay amplitude from
new physics   is negligible.   With $\Delta m_{B_d}(\rm SM)=0.506$ ps$^{-1}$, $\Delta m_{B_s}(\rm SM)=17.80$ ps$^{-1}$, $\beta_d=0.38\pm 0.01$ \cite{Bona:2009tn}, $\beta_s\approx -0.019$ \cite{Chen:2008ug}, $S^{\rm Exp}_{J/\Psi K_S}=0.655\pm 0.024$, $S^{\rm Exp}_{J/\psi \phi} \in (-0.995, -0.285)$ \cite{Barberio:2008fa} and 
$(\Delta m_{B_q})^{\rm Exp}$,
we derive
 \be
-\sin(\phi^{\rm NP}_{d} + \phi^{\rm SM}_d) < 0.2 \,, \non \\
-\sin(\phi^{\rm NP}_{s} + \phi^{\rm SM}_s) <0.985\,. \label{eq:sin_s}
 \ed
In Figs.~\ref{fig:limit_d}(a) and \ref{fig:limit_s}(a),
we present $\Delta m_{B_q}$ and $S_{J/\Psi M}$ with 2$\sigma$ errors of the data as functions of $r_q$ and $\theta^{\rm NP}_{q}$, while the contours for $\sin(\phi^{\rm NP}_{q} + \phi^{\rm SM}_q)$ are displayed in Figs.~\ref{fig:limit_d}(b) and \ref{fig:limit_s}(b)
for $q=d$ and $s$, respectively.
In Fig.~\ref{fig:limit_d}(b), the scattered patten denotes the combined constraints from data of $\Delta m_{B_d}$ and $S_{J/\Psi K_S}$. If we take the central values in Eq.~(\ref{eq:aqsl_value}) as inputs, we immediately obtain that 
$-a^d_{s\ell} < 1.05 \times 10^{-3}$ and $-a^{s}_{s\ell} < 5.32\times 10^{-3}$.  
Although the central values of $|a^{d}_{s\ell}(\rm Exp)|$ and $|a^{s}_{s\ell}(\rm Extr)|$ are larger than our phenomenological analysis, both results are still consistent with each other when  the errors of the data are taken into account.
As a result,  we obtain the  model-independent (MI) result on
the negative like-sign charge asymmetry, given by
 \be
-A^b_{s\ell}(MI) =  -0.506(43) a^d_{s\ell}(MI) - 0.494(43) a^s_{s\ell}(MI) < 3.16 \times 10^{-3}\,.
 \label{MI}
 \ed
 Although the value in Eq.~(\ref{MI}) is smaller
 than the measured value  by   D{\O} in Eq.~(\ref{eq:d0}), it is still one order of magnitude larger than the SM prediction.
 Therefore, $A^b_{s\ell}$ could be a good candidate to probe the new CP violating source in the $B_{d,s}$ systems at Tevatron, LHCb and super-B factories.

\section{ Vector-like quark model 
}\label{sec:model}

\subsection{Z-mediated FCNCs}

By extending the SM with including the new $SU(2)_{L}$ singlet down quarks of
$D_L$ and $D_R$, the extended Yukawa sector becomes
 \be
 -{\cal L}_{Y} &=& \bar Q_L Y_D H d_R + h_D \bar Q_L  H D_R + m_D \bar D_L D_R + h.c.\,,
 \ed
where we have suppressed the flavor indices, $Q_L$ ($H$) is the SU(2) quark (Higgs) doublet, $Y_D$ and $h_D$ are  Yukawa couplings and $m_D$ is the mass of the exotic quark before the electroweak symmetry breaking. When the Higgs field develops the vacuum expectation value (VEV), the mass matrix of the down type quark is given by
 \be
 m_d &=& \left(
           \begin{array}{ccc}
             Y^{ij}_D & | &{\rm 0} \\
             ---& - &- \\
             h^j_D & |& m_D\\
           \end{array}
         \right)\,.
 \ed
Introducing two unitary matrices, the mass matrix can be diagonalized by
 \be
 m^{\rm dia}_{d} &=& V^{L}_{D} m_d V^{R\dagger}_{D}\,. \label{eq:mass}
 \ed

In the SM, since
the interactions of $Z$-boson to fermions are flavor blind,  the flavor in the process with 
the exchange of $Z$-boson is naturally conserved at tree level. In the VQM, the new left-handed quark is an
$SU(2)_L$ singlet and carries the same hypercharge as the right-handed down-type quarks.
The gauge interactions of the left-handed down-type quarks with $Z$-boson are given by
 \be
{\cal L}_{Z}&=& - \frac{gc^{f}_{L}}{2\cos\theta_W} \bar F \ga^{\mu}
X_{F}  P_{L}
  F Z_{\mu}  \label{eq:NC}\,,\non\\
X_{F}&=& \left[
            \begin{array}{ccc}
                \begin{array}{c}
                  \openone_{3\times 3} \\
                \end{array}
              & | & {\bf 0}_{3\times 1} \\
             -\ -\ - & - &\  - \\
                              {\bf 0}_{1\times 3} & |& \xi_D \\
            \end{array}
          \right]\,,
 \ed
where $g$ is the coupling constant of $SU(2)_{L}$, $\theta_W$ is the
Weinberg's angle, $P_{R(L)}=(1\pm\ga_5)/2$,
$F^{T}=(d,s,b,D)$ represents the down-type quarks including the new singlet, $c^{f}_{L}$ is
defined as $c^{f}_{L(R)} = c^{f}_{V} \pm c^{f}_{A}$ with
 \be
 c^{f}_V &=& I^3_f -2\sin^2\theta_W Q_{f}\,,\ \ \ c^{f}_{A}=
 I^3_f \label{eq:cv_ca}
  \ed
in which $I^3_f$ and $Q_f$ are the third component of  the weak
isospin and the electric charge of the particle, respectively, and
$\xi_{f}=-2\sin^2\theta_W Q_f/c^{f}_{L}$. Due to $X_F\neq \openone_{4\times 4}$, accordingly, Eq.~(\ref{eq:NC}) leads to FCNCs at tree level. Since $D_R$ and $q_{R}=(d,s,b)_R$  have the same quantum number, the right-handed quarks are FCNC free at tree level. Following Eq.~(\ref{eq:mass}),  the couplings of  $Z$-boson to fermions in the mass eigenstates
are written by
 \begin{eqnarray}
{\cal L}_{Z}&=& - \frac{gc^{f}_{L}}{2\cos\theta_W} \bar F \ga^{\mu}
\left(V^{L}_{D} X_{F} V^{L^\dagger}_{D}\right) P_{L}
  F Z_{\mu} \,.
 \end{eqnarray}
The FCNC effects could be further formulated as
 \be
 \left(V^{L}_{D} X_{F} V^{L^\dagger}_{D}\right)_{f' f} &=& \delta_{f' f} + (V^{L}_{D})_{f' D} (\xi_D -1) (V^{L^*}_{D})_{fD} = \delta_{f' f} + \lambda_{f' f}\,. \label{eq:lam_ff}
 \ed
Thus, the interaction for $b$-$q$-$Z$ is given by
 \be
 {\cal L}_{b\to q}=-\frac{g c^{d}_{L}\lambda_{qb} }{2\cos\theta_W} \bar q \ga^{\mu}
  P_L b Z_{\mu} + h.c.
 \label{eq:bsint}
 \ed
with 
 \be
 \lambda_{qb}=(\xi_D -1)(V^{L}_D)_{qD}
(V^{L}_D)^*_{bD}\equiv|\lambda_{qb}|\exp\left(i\theta^{\rm
Z}_{q}\right)\,. \non
 \ed
 Clearly, the new free parameters are only $\lambda_{db}$ and $\lambda_{sb}$. When $\lambda_{qb}$ is fixed by the 
 current data, one may have some solid predictions for the relevant processes.

\subsection{ $B_q-\bar B_q$ mixing }

With Eq.~(\ref{eq:bsint}) and the hadronic transition matrix element defined by
 \be
\langle   B_q| \bar q \gamma_{\mu} P_{L(R)} b \bar q \gamma_{\mu} P_{L(R)} b | \bar B_q \rangle
= \frac{1}{3} m_{B_q} f^2_{B_q} \hat B_q \,,
 \ed
the matrix element for  $\bar B_q \to B_q$ mediated by the $Z$-boson at tree level is
obtained as
 \be
M^q_{12}(Z)&=& \frac{G_F \left(\lambda_{qb} c^{d}_{L}\right)^2}{3\sqrt{2}}
m_{B_q} f^2_{B_q} \hat{B_q}=|M^{ q}_{12}(Z)| e^{2i\theta^{ Z}_{q}}\,.
 \ed
In addition to the tree effects, the Z-mediated box and penguin diagrams will induce important linear term in $\lambda_{qb}$ and  it is given by \cite{Barenboim:1997qx}
 \be
 M^{q}_{12}(Loop) &=& -1.3\lambda_{qb} V^*_{tq} V_{tb}\,.
 \ed
Following Eq.~(\ref{eq:m12_SM_NP}), the combination of  the SM and $Z$-mediated tree, box and penguin contributions for the $B_q-\bar B_q$ mixing is given by
 \be
M^q_{12}&=& M^{q}_{12}( SM) + M^{q}_{12}( Z) +M^{q}_{12}(Loop)\non \\
 &=& M^{q}_{12}( SM) R^Z_q e^{2i\beta^{Z}_{q}}\,,
 \ed
where the corresponding parameters in Eq.~(\ref{eq:para}) could be obtained by the following replacements: 
$M^q_{12}(Z)+M^q_{12}(Loop)=M^q_{12}(NP)$, $R^Z_q=R_q (r^Z_q,\theta^Z_q)$ and $\phi^{NP}_q=\phi^{Z}_{q}=2\beta^Z_q$. Hence, the mixing parameter for 
the $B_q$ oscillation is $\Delta m_{B_q}=2|M^q_{12}(SM)| R^Z_q$= $ R^Z_q \Delta m_{B_q}(SM)$.

\subsection{Mixing-induced CP asymmetries}

After deriving  $M^q_{12}$, we now can study the mixing-induced CPAs. The first types of CPAs are the wrong 
and like-sign charged asymmetries,
 defined in Eqs.~(\ref{eq:aqsl}) and  (\ref{eq:Absl}). Since the relationship between 
 the wrong and like-sign asymmetries has been given in Eq.~(\ref{eq:Absl}), we simply formulate the $Z$-mediated $a^q_{s\ell}$ as
 \be
a^q_{s\ell} &=& Im\left(\frac{\Gamma^q_{12}}{M^q_{12}} \right)\,, \non \\
&\approx& \frac{\Delta\Gamma^q(SM)}{\Delta m_{B_q}(SM) \cos\phi^{SM}_{q}} \frac{\sin(\phi^{SM}_{q} + \phi^Z_{q})}{R^Z_q}\,, \label{eq:aqsl_Z}
 \ed
where all SM related quantities are taken to be known.  
Note that  $a^q_{s\ell}$ involves two free parameters, i.e. $|\lambda_{qb}|$ and $\theta^Z_{q}$.

Another type of  the time-dependent CPA is associated with the definite CP  in the final state, defined  by \cite{PDG08}
 \be
A_{f_{CP}}(t)&=& \frac{\Gamma(\bar B_q(t) \to f_{CP})- \Gamma( B_q(t) \to f_{CP})}{\Gamma(\bar B_q(t) \to f_{CP})
+\Gamma( B_q(t) \to f_{CP})}\,,\non \\
&=& S_{f_{CP}} \sin\Delta m_{B_q} t - C_{f_{CP}} \cos\Delta m_{B_q} t\,, \non\\
S_{f_{CP}} &=& \frac{2Im\lambda_{f_{CP}} }{1+|\lambda_{f_{CP}}|^2}\,,\ \ \ C_{f_{CP}}
= \frac{1-|\lambda_{f_{CP}}|^2}{1+|\lambda_{f_{CP}}|^2}\,,
\label{eq:Sf}
 \ed
with
 \be
 \lambda_{f_{CP}} &=&-\left(\frac{M^{B_q^*}_{12}}{M^{B_q}_{12}}\right)^{1/2}
  \frac{A(\bar B\to f_{CP})}{A(B\to f_{CP})} =-  e^{-2i(\beta_q +
  \phi^{\rm NP}_{q}) }\frac{\bar A_{f_{CP}}}{A_{f_{CP}}}\,, \label{eq:lambdaf}
 \ed
where $f_{CP}$ denotes the final CP eigenstate,
$S_{f_{CP}}$ and $C_{f_{CP}}$ are the so-called mixing-induced and direct CPAs, $A_{f_{CP}}$
and $\bar A_{f_{CP}}$ are the amplitudes of $B$ and $\bar B$ mesons decaying
to $f_{CP}$ and $\bar A_{f_{CP}}/A_{f_{CP}}=-\eta_{f_{CP}} A_{f_{CP}}(\theta_W\to -\theta_W)/A_{f_{CP}}
(\theta_W)$ with $\eta_{f_{CP}}$ and $\theta_W$ being the CP eigenvalue of $f_{CP}$ and
the weak CP phase, respectively. Clearly, besides the phase  in the $\Delta B=2$ process, the mixing-induced CPA is also related to
the phase in the $\Delta B=1$ process. Due to $B\to \eta' K_S$ involving more complicated and uncertain QCD effects, in this paper, we will concentrate on $f_{CP}=J/\Psi K_S$ and $\phi K_S$ for $q=d$ and  $f_{CP}=J/\Psi \phi$ for $q=s$.

For $\Delta B=1$ processes, we also need to know the flavor conserving interactions. 
The couplings of $Z$-boson to fermions in the SM are summarized as
 \be
{\cal L}^{SM}_Z&=&-\frac{g}{2\cos\theta_W} \sum _{f} \bar f \ga^\mu\left( c^f_V  - c^f_A \ga_5\right) f Z_\mu\,, \label{eq:lang_Z}
\ed
where $f$ denotes any fermions and $c^f_{V(A)}$ is given in Eq.~(\ref{eq:cv_ca}). 
Using Eqs.~(\ref{eq:bsint}) and (\ref{eq:lang_Z}), the $Z$-mediated Hamiltonian for $b\to q q' \bar q'$ decays is obtained by
 \be
{\cal H}^Z_{b\to q q' \bar q'}&=& \frac{G_F}{\sqrt{2}} \left( \frac{\lambda_{qb}c^d_L}{2}\right) (\bar q b)_{V-A} \sum _{q'=u,d,s,c} \left( c^{q'}_{L}(\bar q' q')_{V-A} + c^{q'}_{R} (\bar q' q')_{V+A} \right) 
 \label{eq:bs_int}
 \ed
where $(\bar f' f)_{V\pm A}=\bar f' \ga^\mu (1\pm \ga_5) f$. Clearly, the $Z$-mediated effects for $b\to q q'\bar q'$ are similar to 
the standard electroweak penguins but the SM contributions are small. Since $B_d\to J/\Psi  K_S $ and $B_s\to J/\Psi \phi$ decays are dominated by the tree diagrams, the penguin-like effects can be regarded to be relatively small and insignificant in the $b\to s c\bar c$ processes. On the contrary, since $b\to ss\bar s$ is a penguin dominant process, the $Z$-mediated effects  are naturally comparable with the SM contributions. Hence, we will only focus on $B\to \phi K_S$. In order to deal with the hadronic effects in nonleptonic $B_q$ decays, we employ the naive factorization approach (NFA). 
The decay amplitude combined the SM with $Z$-mediated contributions for $B\to \phi K$ is written as
 \be
\bar A_{\phi \bar K^0}&=& \la \phi \bar K| {\cal H}_{b\to ss\bar s}|\bar B^0 \ra\,, \non\\
&=& \frac{G_F}{\sqrt{2}} V^*_{ts} V_{tb} (a^{\rm SM} + a^Z_{s}) \la \phi|\bar s \ga_\mu s|0\ra \la \bar K^0| \bar s \ga^\mu b|\bar B\ra \,,
 \ed
with
 \be
 a^{\rm SM}_s&=&a_3+ a_4 + a_5\,,\non\\
 a_3&=&C_3 +\frac{C_4}{N_C}\,, \ \ \ a_4 = C_4 + \frac{C_3}{N_C}\,, \non \\
 a_5 &=& C_5 + \frac{C_6}{N_C}\,, \ \ \ a^Z_s= -\frac{\lambda_{sb}c^d_L }{V^*_{ts} V_{tb}} \left(c^s_V+\frac{c^s_L}{2N_C} \right) \,,\non
  \ed
where
 $N_C$ is the number of colors and $C_{3-6}$ are the effective Wilson coefficients from the
gluon penguins of the SM \cite{BBL}. We note that the
electroweak penguin contributions are very small and neglected in the analysis. 

Consequently, the ratio of amplitudes for $\bar B_d\to \phi K_S$ and $B_d\to \phi K_S$ decays is written as 
 \be
\frac{\bar A_{\phi K_S}}{A_{\phi K_S} }= -e^{2i\beta_s} \frac{a^{SM} + a^{Z}_{s}}{a^{SM} + a^{Z^*}_{s}}=-e^{2i(\beta_s
 + \delta^{ Z}_{s})}  \ed
with
 \be
\tan\delta^{Z}_{s} &=& \frac{|a^Z_{s}| \sin(\theta^{Z}_{s}-\beta_s) }{a^{SM} + |a^{Z}_{s}| \cos(\theta^{Z}_{s} -\beta_s)}\,. \non
 \ed
  From Eqs.~(\ref{eq:Sf}) and (\ref{eq:lambdaf}), the mixing-induced CPA through the $\phi K_S$ mode is obtained as
 \be
 S_{\phi K_S} &\equiv& \sin2\beta_{\phi K_S} = \sin2(\beta_d + \beta^{Z}_{d}-\beta_s
 -\delta^{Z}_{s})
 \,.
 \label{eq:S_phik}
 \ed
Similarly,  the CPAs through $ J/\Psi (K_S, \phi)$ channels are simply given by
 \be
 S_{J/\Psi K_S}&\equiv& \sin2\beta_{J/\Psi K_S} \approx  \sin(2\beta_d + \phi^{Z}_{d})\,, \non \\
 S_{J/\Psi \phi} &\equiv& \sin2\beta^{J/\Psi \phi}_{s} \approx \sin(2\beta_s + \phi^{Z}_{s})\,. \label{eq:Sjpsi_phi}
 \ed
Although  $\sin2\beta_{J/\Psi K_S}$ has been measured at a precision level, it might be difficult to confirm whether  new physics exists by observing $\sin2\beta_{J/\Psi
K_S}$ alone. Nevertheless, one can investigate
 a new asymmetry, defined by \cite{Grossman:1996ke}
 \be
 \Delta S_{\beta_d}= \sin2\beta_{J/\Psi K_S} -\sin2\beta_{\phi K_S}\,, \label{eq:dbeta}
 \ed
in which the SM prediction is less than around $5\%$ \cite{Grossman:1996ke}. Clearly, if a large value of
 $\Delta S_{\beta_d}$ is measured, it will be a strong hint for new physics beyond the SM.

\subsection{ $b\to q \ell^+ \ell^-$ and $B_q\to \mu^+ \mu^-$ decays}

In addition to the CP violating observables, the other interesting environment to probe the new physics effects is rare decays in which the predicted branching ratios (BRs) in the SM are small. Although the BR is not a direct CP violating observable,
 it is still sensitive to the CP violating effect via the squared imaginary coupling.  In most exclusive decay processes, the BRs are associated with uncertain nonperturbative hadronic effects.
 To reduce the QCD uncertainties, we choose inclusive $b\to q \ell^+ \ell^-$ and exclusive $B_q\to \ell^+ \ell^-$ decays
 as the candidates to probe the new physics effects, where the hadronic  effects could be  controlled well.  

Using Eqs.~(\ref{eq:bsint}) and (\ref{eq:lang_Z}), the effective Hamiltonian for $b\to q \ell^+ \ell^-$ mediated by $Z$-boson is found 
to be
 \be
{\cal H}^Z_{b\to q \ell^+ \ell^-} &=& \frac{G_F}{\sqrt{2}} \lambda_{qb} c^d_L (\bar q b)_{V-A} \left[c^\ell_V (\bar\ell \ell)_V - c^\ell_A (\bar \ell \ell)_A \right]
 \ed
where $(\bar\ell \ell)_V=\bar \ell \ga^\mu \ell$ and $(\bar\ell \ell)_A=\bar\ell \ga^\mu \ga_5 \ell$. Combining with the SM contributions, the decay amplitude for $b\to q\ell^+ \ell^-$ is written as
 \be
{\cal H}_{b\to q\ell^+ \ell^-} &=& -\frac{G_F \alpha}{\sqrt{2} \pi} V^*_{tq} V_{tb}
\left[ \left(C_{9} \bar q \ga_\mu P_L b -\frac{2m_b}{k^2} C^{SM}_{7\ga} \bar q i\sigma_{\mu\nu} k^\nu P_R b \right) \bar \ell \ga^\mu \ell \right. \non \\
&+& \left. C_{10} \bar q \ga_\mu P_L b \bar \ell \ga_\mu \ga_5 \ell
\right] \label{eq:int_bqll}
 \ed
where $k_\mu=(p_{\ell^+} + p_{\ell^-})_\mu$, $k^2$ is the invariant mass of the lepton pair, and
 \be
C_9 &=& C^{SM}_9(m_b) - \frac{2\pi}{\alpha}\frac{\lambda_{qb} c^d_L c^\ell_V}{V^{*}_{tq} V_{tb}}\,, \non \\
C_{10} &=& C^{SM}_{10} +  \frac{2\pi}{\alpha}\frac{\lambda_{qb} c^d_L c^\ell_A}{V^{*}_{tq} V_{tb}}\,.
 \ed
The explicit expressions of $C^{SM}_{9, 10}$ could be found in Ref.~\cite{BBL}.
Accordingly, the differential decay rate is  \cite{BBL}
 \be
\frac{d\Gamma(b\to q \ell^+ \ell^-)}{d\hat s} &=& \Gamma(b\to c e\bar \nu_e) \frac{|V^*_{ts} |^2}{|V_{cb}|^2}\frac{\alpha^2}{4\pi^2} \frac{(1-\hat s)^2}{f(z) k(z)} \non \\
&\times& \left[ (1+2\hat s) \left(|C_9|^2 + |C_{10}|^2 \right) + 4\left( 1+\frac{2}{\hat s}\right) |C^{SM}_{7\ga}|^2 + 12 C^{SM}_7 Re C_9
\right]\,, \non \\
f(z)&=&1-8z^2+8z^6 -z^8 -24 z^4 \ln z\,, \non \\
k(z)&=& 1-\frac{2\alpha_s}{3\pi} \left[\left(\pi^2 - \frac{31}{4}\right) (1-z)^2 + \frac{3}{2} \right]\,,
 \ed
where $\hat s=k^2/m^2_b$,  $z=m_c/m_b$ and $\Gamma(b\to c e \bar \nu _e)$ is used to cancel the uncertainties from the CKM matrix elements and $m^5_b$. Moreover, with Eq.~(\ref{eq:int_bqll}) and the $B_q$ meson decay constant, defined by
 \be
\la 0| \bar q \ga^\mu \ga_5 b|\bar B_q\ra &=& i f_{B_q} p^\mu_{B_q}\,, 
 \ed
the BR for  $B_q\to \ell^{+} \ell^{-}$ is straightforwardly obtained by
 \be
{\cal B}(B_q\to \ell^+ \ell^-)={\cal B}^{SM}( B_q \to \ell^+ \ell^-)  \left| 1-\frac{\pi}{\alpha}\frac{\lambda_{qb} c^d_L}{V^*_{tq} V_{tb} C^{SM}_{10}}\right|^2, \label{eq:brBsmumu}
\ed
where
 \be
{\cal B}^{SM}( B_q \to \ell^+ \ell^-) &=& \tau_{B_q}\frac{G^2_F \alpha^2}{16\pi^3} |V^*_{tq}  V_{tb}|^2
m_{B_q}f^2_{B_q}m^2_{\ell}|C^{SM}_{10}|^2 \left(
1-\frac{4m^2_{\ell}}{m^2_{B_q}}\right)^{1/2}
\,. \non
%
 \ed

\section{Numerical analysis}\label{sec:num}

As stated earlier, there are four new unknown parameters for $b\to (d, s)$ transitions in the VQM, i.e., $|\lambda_{db,sb}|$ and $\theta^{Z}_{d,s}$. Although we have  model-independently shown the possible severe constraints in Sec.~\ref{sec:wrong-like},
 in a specific model, we have to consider more relevant bounds. As $\lambda_{f' f}= (\xi_D-1)(V^L_D)^*_{f' D} (V^L_D)_{fD} $ defined in Eq.~(\ref{eq:lam_ff}),  the
 $s\to d$ transition is associated with $(V^L_D)^*_{d D} (V^L_D)_{sD} $ while the $b\to (d, s)$ ones 
 depend on $(V^L_D)^*_{d D} (V^L_D)_{bD} $ and $(V^L_D)^*_{s D} (V^L_D)_{bD} $, respectively.
  Thus, $(V^L_D)^*_{d D}$ and $ (V^L_D)^*_{s D}$ appearing in $b\to (d,s)$ also occur in $s\to d$. We see clearly that $K^0-\bar K^0$  and $B_{d,s}-\bar B_{d,s}$ mixings are strongly correlated. Since $\Delta m_{K}$ and the indirect (direct) CP violating parameters denoted by $\epsilon_{K}$ ($\epsilon'_K$) are much smaller than those in the $B_q$ systems, the stringent constraints could not make $\lambda_{db}$ and $\lambda_{sb}$ be large simultaneously. Moreover, by the results in Sec.~\ref{sec:wrong-like}, we know that $\Delta m_{B_d}$ and $\sin2\beta_{J/\Psi K_S}$ will push the allowed parameter space of $\lambda_{db}$ to the region with small values. 
  Without loss of generality, for simplicity we directly set the effects of $\lambda_{db}$ be insignificant and ignorable. Hence, we will focus on the contributions of $\lambda_{sb}$ in our numerical presentation, which relate to various $b\to s$  processes.

For numerical calculations and constraints, we list the useful values in Table~\ref{tab:inputs},  where  the relevant CKM matrix elements $V_{td}=|V_{td}|\exp(-i\beta_d)$ and $V_{ts}=-|V_{ts}|\exp(-i\beta_s)$ are obtained from the UTfit Collaboration~\cite{Bona:2009tn}, the decay constant of $B_q$ is referred to the result given  by the HPQCD Collaboration~\cite{Gamiz:2009ku}, the CDF and D$\O$ average value
of $\Delta m_{B_s}$ is from Ref.~\cite{Barberio:2008fa} and the SM
Wilson coefficients for $b\to q q' \bar q'$ and $b\to q\ell^+ \ell^-$  are obtained from Ref.~\cite{BBL}. The upper limit for ${\cal B}(B_s\to \mu^+ \mu^-)$ with $95\%$ confidence level (C. L.) is quoted by the latest D{\O}  measurement \cite{Abazov:2010fs}. Other inputs are from the particle data group (PDG) \cite{PDG08}.
\begin{table}[hptb]
\caption{Experimental data and numerical inputs for the parameters in the SM.
 } \label{tab:inputs}
\begin{ruledtabular}
\begin{tabular}{ccccc}
  $|V_{td}|$ & $\beta_d$ & $ |V_{ts}|$ & $\beta_s$ & $m_{B_d}$ 
 \\ \hline
 $8.51(22)\times 10^{-3}$ & $(22\pm 0.8)^{\circ}$ & $ 4.07(22)\times 10^{-2}
 $ & $-(1.03\pm 0.06)^{\circ}$ & 5.28 GeV  \\ \hline\hline
  $m_{B_s}$ & $f_{B_d} \sqrt{\hat B}_d$ & $f_{B_s} \sqrt{\hat{B_s}}$ & $f_{B_d}$ & $f_{B_s}$
   \\ \hline
  5.37 GeV & $(216\pm 15)$ MeV & $(266\pm 18)$ MeV & $190\pm 13 $ MeV & $231 \pm 15$ MeV   \\ \hline\hline
 $S^{\rm Exp}_{J/\Psi K_S}$& $S^{\rm Exp}_{\phi K_S}$  & $(\Delta m_{B_d})^{\rm Exp}$ & $(\Delta m_{B_s})^{\rm Exp}$ & ${\cal B}^{\rm Exp}(b\to s\ell^+ \ell^-)$     \\ \hline
 $0.655 \pm 0.024$ & $0.44^{+0.17}_{-0.18}$ & $0.507 \pm 0.005$ ps$^{-1}$ & $17.77 \pm 0.12$ ps$^{-1}$ & $(4.5 \pm 1.0)\times 10^{-6}$   \\ \hline\hline
 ${\cal B}^{\rm Exp}(B_s\to \mu^+ \mu^{-})$   & $C_3$  & $C_4$ & $C_5$   & $C_6$  \\ \hline
 $<5.1 \times 10^{-8}$   & $0.013$ & $-0.0335$ & $0.0095$ & $-0.0399$ \\ \hline
  $C^{SM}_{7\ga}$  & $C^{SM}_9$ & $C^{SM}_{10}$ &   $\sin^2\theta_W$ & $\alpha(m_Z)$ \\ \hline
  $-0.305$ & $4.344$ & $-4.430$ &  0.231& 1/129  \\
 
\end{tabular}
\end{ruledtabular}
\end{table}

\begin{figure}[hpbt]
\includegraphics*[width=4. in]{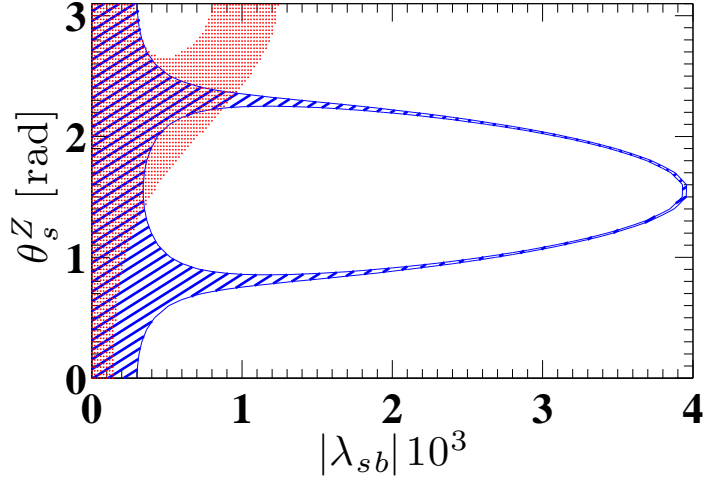}
\caption{  Constraint of $|\lambda_{sb}|$ and $\theta^Z_{s}$ from $\Delta m_{B_s}$ and ${\cal B}(b\to s \ell^+ \ell^-)$.  }
 \label{fig:constraint_Z}
\end{figure}

Before we discuss the VQM predictions, it is necessary to know which processes involve less  hadronic uncertainties and could give the strict constraints. We find that in addition to  $\Delta m_{B_s}$,  the observed inclusive $b\to s\ell^+ \ell^-$ decays with $\ell=e, \mu$ are the good candidates. Although  the possible constraint of $\sin2\beta^{J/\Psi \phi}_s$ has been mentioned in Sec.~\ref{sec:wrong-like}, as shown in Eq.~(\ref{eq:sin_s}),  its current measurement cannot provide any significant bound. We present 
 $\Delta m_{B_s}$  (down-left hatched) and ${\cal B}(b\to s\ell^+ \ell^+)$ (dotted) with $2\sigma$ errors of the data as  functions of $|\lambda_{sb}|$ and $\theta^Z_s$  in Fig.~\ref{fig:constraint_Z}, in which $|\lambda_{sb}|$ is in units of $10^{-3}$.  
  From the figure, we see  that ${\cal B}(b\to s\ell^+ \ell^-)$ further limits the upper value of $|\lambda_{sb}|$ to be around $10^{-3}$.   In general, the range of the CP violating phase $\theta^{Z}_s$ is $[-\pi, \pi]$.   For simplicity, we just show the results within $[0, \pi]$. The pattern of the constraint in $[-\pi,0]$ is similar to that in $[0, \pi]$.

\begin{figure}[hpbt]
\includegraphics*[width=5. in]{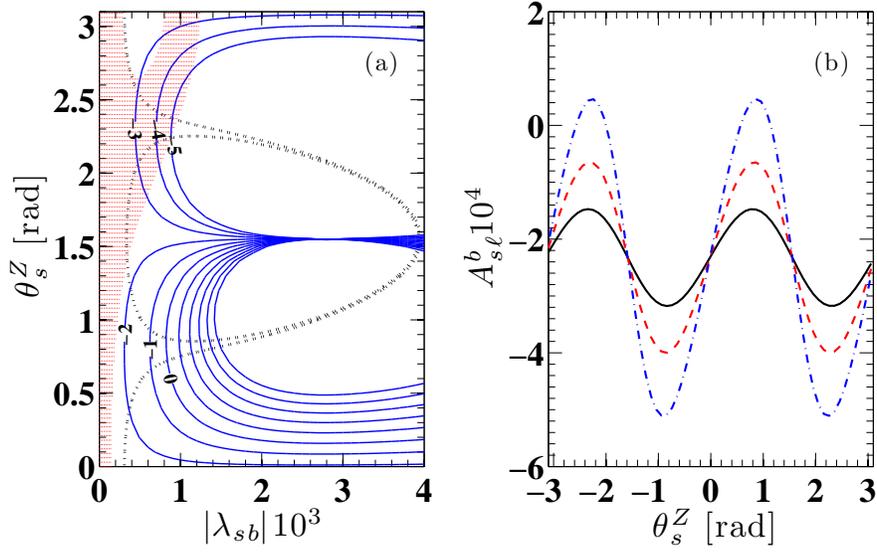}
\caption{  (a) Contours for $A^b_{s\ell}$ (in units of $10^{-4}$)
 as a function of $|\lambda_{sb}|$ and  $\theta^Z_{s}$
 and
 (b) $A^b_{s\ell}$ as a function of $\theta^Z_s$, where 
  the sold, dashed and dash-dotted lines represent $|\lambda_{sb}|=(0.5, 0.7, 0.9)\times 10^{-3}$, respectively. }
 \label{fig:absl_Z}
\end{figure}

Since we set the $Z$-mediated $b\to d$ transition be negligible, the wrong-sign charge asymmetry for  $B_d$ decays is  ascribed to the SM contribution. We take $a^d_{s\ell}(SM)=-4.8\times 10^{-4}$ for our numerical estimates. Using Eq.~(\ref{eq:aqsl_Z}) for $a^s_{s\ell}$ and Eq.~(\ref{eq:Absl}) for the like-sign charged asymmetry,  the contours for $A^{b}_{s\ell}$ as a function of $|\lambda_{sb}|$ and $\theta^Z_s$ are shown in Fig.~\ref{fig:absl_Z}(a), where the numbers in the plot are units of $10^{-4}$. We also plot  $A^{b}_{s\ell}$ as a function of $\theta^Z_s$ with fixing $|\lambda_{sb}|$ in Fig.~\ref{fig:absl_Z}(b), in which the solid, dashed and dash-dotted lines denote $|\lambda_{sb}|=(0.5, 0.7, 0.9)\times 10^{-3}$, respectively. 
From  Fig.~\ref{fig:absl_Z}, we see  that due to the constraint of ${\cal B}(b\to s \ell^+ \ell^-)$, 
the absolute value of the like-sign charge asymmetry $A^{b}_{s\ell}$
can be as large as
$5\times 10^{-4}$.  Although the result is not enhanced by order of magnitude, 
it could be still a factor of two larger than the SM prediction. 

\begin{figure}[bthp]
\includegraphics*[width=5. in]{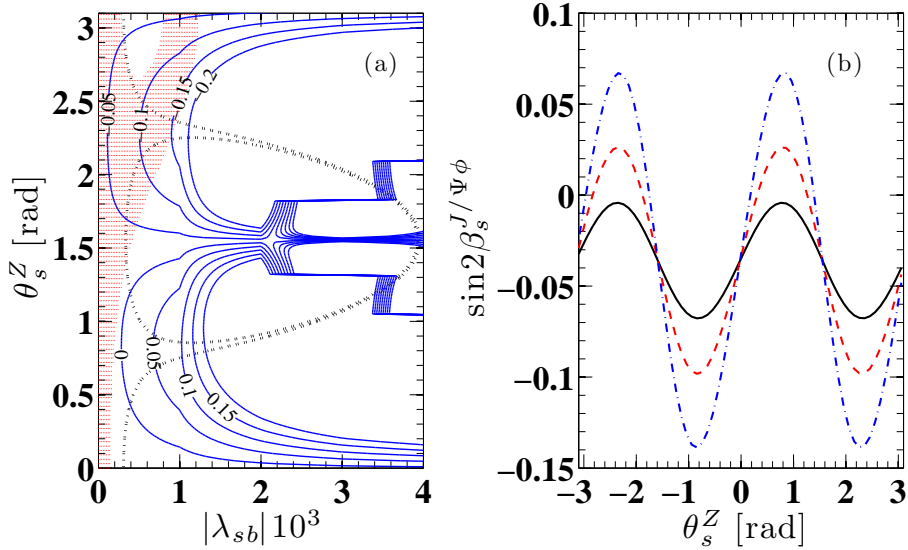}
\caption{  (a) Contours for $\sin2\beta^{J/\Psi \phi}_s$ as a function of $|\lambda_{sb}|$ and  $\theta^Z_{s}$ and
 (b) $\sin2\beta^{J/\Psi \phi}_s$ as a function of $\theta^Z_s$ with the same legend as Fig.~\ref{fig:absl_Z}(b).
 }
 \label{fig:Jpsi_Phi_Z}
\end{figure}

Next, we analyze the time-dependent CPA in the $B_s\to J/\Psi \phi$ decay. 
When Z-mediated $b\to d$ effects are neglected, it is easy to find that $A^{b}_{s\ell}$ and $S_{J/\Psi
\phi}$ defined in Eq.~(\ref{eq:Sf}) have a strong correlation. By using Eq.~(\ref{eq:Sjpsi_phi}),  the contours for the time-dependent CPA of $\sin2\beta^{J\Psi \phi}_s$ as a function of $|\lambda_{sb}|$ and $\theta^Z_s$ are displayed in Fig.~\ref{fig:Jpsi_Phi_Z}(a). Moreover, $\sin2\beta^{J\Psi \phi}_s$ as a function of $\theta^Z_s$ with fixing $|\lambda_{sb}|$ is shown in Fig.~\ref{fig:Jpsi_Phi_Z}(b) with the same legend as Fig.~\ref{fig:absl_Z}(b).
According to the results, we find that when the constraints of $\Delta m_{B_s}$ and ${\cal B}(b\to s\ell^+ \ell^-)$ are taken into account at the same time, the sign of $\sin2\beta^{J/\Psi \phi}_s$ favors negative, which is the same as that indicated by CDF and D{\O} measurements. Although the upper limit on the magnitude is smaller than the current data, it could
still be $15\%$, whereas the SM prediction is only around $4\%$.

\begin{figure}[bthp]
\includegraphics*[width=5. in]{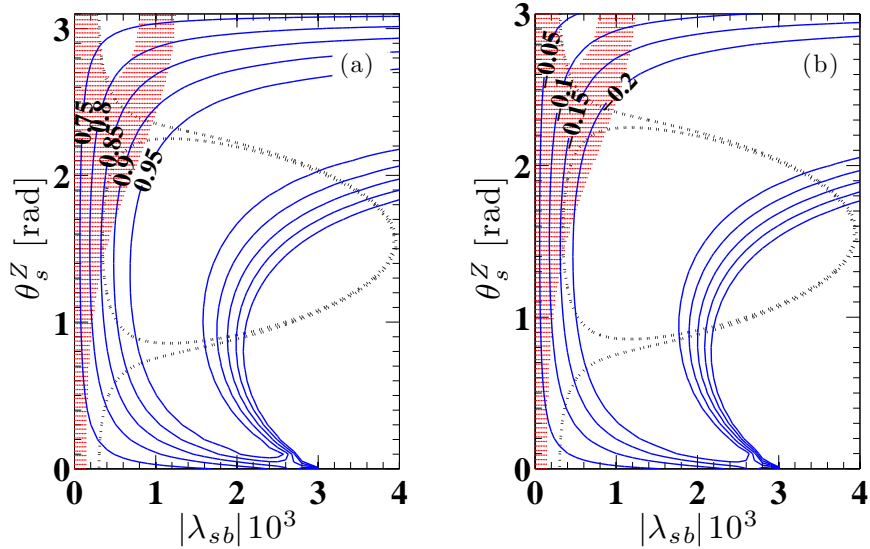}
\caption{  (a) [(b)] Contours for $\sin2\beta_{\phi K_S}$ [$\Delta S_{\beta_d}$] as a function of $|\lambda_{sb}|$ and  $\theta^Z_{s}$.  }
 \label{fig:PhiKS_contour_Z}
\end{figure}

\begin{figure}[bthp]
\includegraphics*[width=5. in]{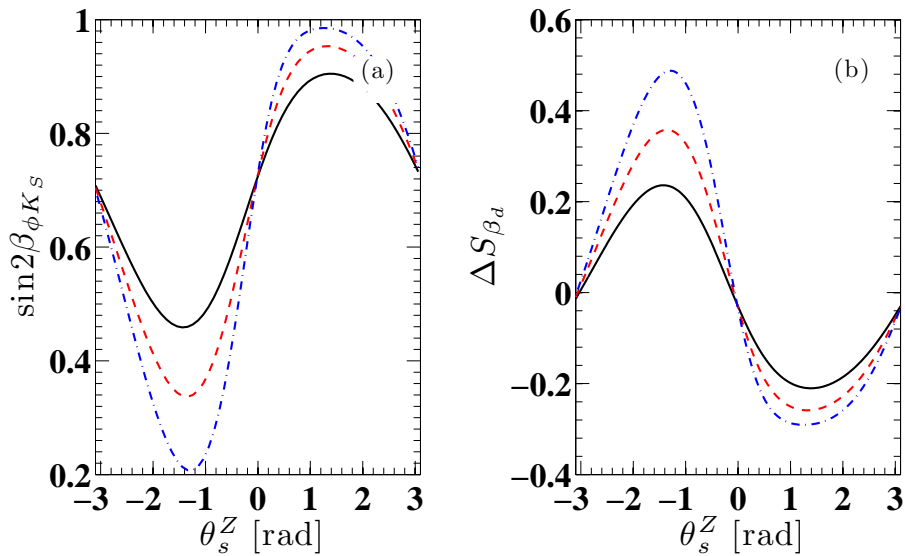}
\caption{  (a) [(b)]  $\sin2\beta_{\phi K_S}$ [$\Delta S_{\beta_d}$] as a function of  $\theta^Z_{s}$
with the same legend as Fig.~\ref{fig:absl_Z}(b).}
 \label{fig:PhiKS_sin_Z}
\end{figure}

In terms of the early analysis,  the penguin-like $Z$-mediated effect for $b\to s c\bar c$  in Eq.~(\ref{eq:bs_int}) could be estimated as
 \be
  \left|\frac{\lambda_{sb}c^d_L }{V^*_{ts} V_{tb}} c^c_V\right| <0.0048 \sim |a_5|. 
 \ed
It is clear that the new effect  to the decay amplitude of $B\to J/\Psi K_S$ is insignificant as the case in the SM. Thus, we have $\sin2\beta_{J/\Psi K_S}\approx \sin2\beta_{J/\Psi K_S}(SM)\approx 0.695$ in the $Z$-mediated VQM.  In order  to probe the new CP violating source arising
 from $SU(2)$ singlet exotic quarks, the best observable  is the time-dependent CPA in the
 $B_d\to \phi K_S$ decay, where $\sin2\beta_{\phi K_S}$ and $\sin2\beta_{J/\Psi K_S}$, defined by Eqs. ~(\ref{eq:S_phik}) and (\ref{eq:Sjpsi_phi}), have similar values in the SM,  respectively. To understand the influence of $Z$-mediated effects on the 
 CPA in $B_d\to \phi K_S$, we display the contours for $\sin2\beta_{\phi K_S}$ as a function of $|\lambda_{sb}|$ and $\theta^Z_s$ in Fig.~\ref{fig:PhiKS_contour_Z}(a).  From the result, we find that $\sin2\beta_{\phi K_S}$ could approach $0.90$  when 
  $|A^b_{s\ell}|$ is a factor of two larger than the SM prediction. Furthermore, in Fig.~\ref{fig:PhiKS_contour_Z}(b) we present the contours for $\Delta S_{\beta_d}$, the difference in the CPA between $J/\Psi K_S$ and $\phi K_S$ modes defined by Eq.~(\ref{eq:dbeta}). Clearly, the difference of $-20\%$ could be achieved. It is interesting to mention that $\sin2\beta_{\phi K_S}$ in the VQM is larger than $\sin2\beta_{J/\Psi K_S}$ in $[0, \pi]$, whereas the situation is reversed in $[-\pi, 0]$. Although the current data in $B_d\to \phi K_S$ prefers  the latter case, in this region $|A^b_{s\ell}|$ is even smaller than the SM result. Due to the current accuracy of the data, it is hard to tell which solution is more close to the reality. Hence, more precise measurements are necessary.  For further comprehending the $\theta^Z_s$ dependence, we plot $\sin2\beta_{\phi K_S}$ and $\Delta S_{\beta_d}$ as functions of $\theta^Z_s$ in Fig.~\ref{fig:PhiKS_sin_Z}(a) and (b), where the solid, dashed and dash-dotted lines denote $|\lambda_{sb}|=(0.5, 0.7, 0.9)\times 10^{-3}$, respectively.

\begin{figure}[bthp]
\includegraphics*[width=4. in]{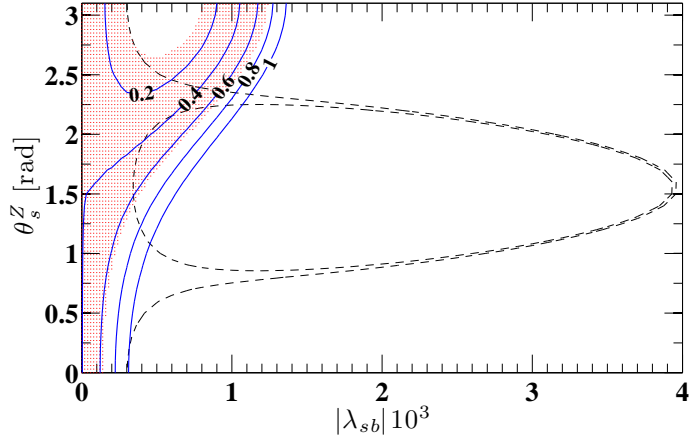}
\caption{  Contours for ${\cal B}(B_s\to \mu^+ \mu^-)$ as a function of $|\lambda_{sb}|$ and $\theta^Z_s$, where the numbers in the plot are in units of $10^{-8}$.  }
 \label{fig:Bsmumu}
\end{figure}

Finally, we analyze the rare decays of $B_q\to \ell^+ \ell^-$. As discussed earlier, the $b\to d$ transition in 
the $Z$-mediated VQM is suppressed and therefore, we will concentrate on $B_s\to \ell^+ \ell^-$.  Since the leptonic 
 process is helicity-suppressed, only the heavier charged leptonic modes are interesting. 
 However,
since the experiments only provide the limits on $B_s\to \mu^+ \mu^-$, we  study the influence of $Z$-mediated effects on the muon channel. Using Eq.~(\ref{eq:brBsmumu}) and the values in the Table~\ref{tab:inputs}, the contours for ${\cal B}(B_s\to \mu^+ \mu^-)$ as a function of $|\lambda_{sb}|$ and $\theta^Z_s$ are displayed in Fig.~\ref{fig:Bsmumu}. We find that the upper value of ${\cal B}(B_s\to \mu^+ \mu^-)$ is around $0.6\times 10^{-8}$ whereas the SM result of ${\cal B}^{SM}(B_s\to \mu^+ \mu^-)$ is around $0.39 \times 10^{-8}$.

\section{Conclusion}\label{sec:conclusion}

We have 
 model-independently
 studied the charge and CP asymmetries as well as FCNCs in the various $B_{d,s}$ processes.
In particular, we have found that
   $(-A^b_{s\ell})< 3.16\times 10^{-3}$ when the constraints from the $B_q-\bar B_q$ mixings and the time-dependent CP asymmetries (CPA) for $B_q\to J/\Psi M_q$ with $M_q=K,\phi$ and $q=d,s$ are taken into account. Although the upper value is smaller than the data of 
the new D{\O} measurement, it is still one order of magnitude larger than the standard model (SM) prediction and sensitive to new CP violating effects. 
We have also explored the VQM to illustrate the possible  large effects on $|A^{b}_{s\ell}|$ and FCNCs in the $B_{d,s}$ processes.
Explicitly, we have shown that (a) the like-sign charge asymmetry could be enhanced by a factor of two in magnitude; (b) the CPA of $\sin2\beta^{J/\Psi \phi}_s$ could reach to $-15\%$; (c) the CPA of $\sin2\beta_{\phi K_S}$ could be higher than $\sin2\beta_{J/\Psi K_S}$ when $|A^b_{s\ell}|$ is larger than the SM prediction; and (d) the BR for $B_s\to \mu^+ \mu^-$ could be as large as $0.6\times 10^{-8}$.

\begin{acknowledgments}
 This work is supported in part by
the National Science Council of R.O.C. under Grant Nos: NSC-97-2112-M-006 -001-MY3, NSC-95-2112-M-007-059-MY3 and NSC-98-2112-M-007-008-MY3.

\end{acknowledgments}

\end{document}